# Electron-phonon anomaly related to charge stripes: static stripe phase versus optimally-doped superconducting $La_{1.85}Sr_{0.15}CuO_4$

D. Reznik[1,2], L. Pintschovius[1], M. Fujita[3], K. Yamada[3], G.D. Gu[4], and J.M. Tranquada[4]

*1 Forschungszentrum Karlsruhe, Institut für Festkörperphysik, P.O.B. 3640, D-76021 Karlsruhe, Germany*

*2 Laboratoire Léon Brillouin, CE Saclay, F-91191 Gif-sur-Yvette Cedex, France*

*3 Institute for Material Research, Tohoku University, Katahira, Aoba-ku, Sendai, 980-8577, Japan.*

*4 Physics Department, Brookhaven National Laboratory, Upton, New York 11973*

Abstract

Inelastic neutron scattering was used to study the Cu-O bond-stretching vibrations in optimally doped $La_{1.85}Sr_{0.15}CuO_4$ ($T_c$ = 35 K) and in two other cuprates showing static stripe order at low temperatures, i.e. $La_{1.48}Nd_{0.4}Sr_{0.12}CuO_4$ and $La_{1.875}Ba_{0.125}CuO_4$. All three compounds exhibit a very similar phonon anomaly, which is not predicted by conventional band theory. It is argued that the phonon anomaly reflects a coupling to charge inhomogeneities in the form of stripes, which remain dynamic in superconducting $La_{1.85}Sr_{0.15}CuO_4$ down to the lowest temperatures. These results show that the phonon effect indicating stripe formation is not restricted to a narrow region of the phase diagram around the so-called 1/8 anomaly but occurs in optimally doped samples as well.

## 1. Introduction

One of the focal points of the debate on high temperature superconductors is whether their electronic state is spatially homogeneous, as in ordinary metals, or inhomogeneous due to self-organization of the charge carriers.[1,2] In particular, it has been argued that holes doped into copper-oxygen planes form stripes, i.e. they segregate into lines, which act as domain walls for an antiferromagnetically ordered background. Static long-range stripe order has been observed only in a few special compounds such as $La_{1.48}Nd_{0.4}Sr_{0.12}CuO_4$ and $La_{1.875}Ba_{0.125}CuO_4$ where a small structural modification at low temperatures provides the anisotropic pinning for the stripes while superconductivity is greatly suppressed.[3,4] Above the ordering temperature, the stripes are assumed to be purely dynamic, which makes their detection extremely difficult.[5] Recently, a giant softening of the phonon whose polarization pattern matches the lattice deformation induced by the charge stripes has been identified in the above-mentioned compounds showing static stripe order[6]. It has been found that this phonon anomaly largely survives to temperatures above the ordering temperature and hence can be considered as a signature of dynamic charge order. Here, we report further inelastic neutron scattering results on the phonon anomaly in nonsuperconducting $La_{1.875}Ba_{0.125}CuO_4$ and, in particular, of our search for a similar effect in optimally doped $La_{1.85}Sr_{0.15}CuO_4$ ($T_c$ = 35 K). We find that a very similar phonon anomaly exists indeed in $La_{1.85}Sr_{0.15}CuO_4$ indicating that compounds with high superconducting transition temperatures show dynamic charge order.

## 2. Experimental Details

The experiments were performed on the 1T triple-axis spectrometer at the ORPHEE reactor at LLB, Saclay. A copper 220 reflection was utilized as the monochromator in order to achieve high resolution, while Pyrolytic Graphite (PG002) was used as analyzer, with the final energy usually held fixed at 14.7meV to allow the use of a PG filter to suppress higher orders. In the following, wavevectors are given in reciprocal lattice units of ($2\pi/a$, $2\pi/b$, $2\pi/c$), where a=b=3.78 Å, and c=13.18 Å. The single crystal sample of $La_{1.875}Ba_{0.125}CuO_4$ (for details see ref. 4) weighing 5g was mounted on the spectrometer with the b-axis vertical for the measurements whose results appear in figure 2 and with the c-axis vertical for other figures. The 5g crystal of $La_{1.85}Sr_{0.15}CuO_4$ always had the c-axis vertical. For temperature control the samples were mounted in a closed-cycle refrigerator. For each compound, the scattering cross section was measured in the Brillouin zone (BZ) near the reciprocal lattice vector **G**=(5 0 0) as shown in Fig. 1. These wave vectors were chosen because they provide the largest accessible dynamic structure factor of the bond-stretching phonons. For neutron momentum transfer $\hbar\mathbf{Q}$, the phonon wave vector is $\mathbf{q} = \mathbf{G} - \mathbf{Q}$. In $La_{1.875}Ba_{0.125}CuO_4$, the charge-stripe modulation appears at $\mathbf{Q}_{co} = \mathbf{G} - \mathbf{q}_{co} = (4.75\ 0\ 0)$.

## 3. Results

Figure 2a illustrates the electron-phonon anomaly at the stripe ordering wavevector in $La_{1.875}Ba_{0.125}CuO_4$ at T = 10K. It is immediately clear that the well-defined phonon branch dispersing downward from the zone center acquires a broad low energy tail around $\mathbf{Q}_{co}$. At slightly larger $\mathbf{q}$ the data can be fit with a two-peak structure, with the peaks merging by $\mathbf{q}=(0.4,0,0)$. Figure 2b demonstrates that the integrated intensity of the entire feature agrees very well with the calculated[7] structure factor of the bond-stretching mode. The broadest lineshape appears at the stripe ordering wavevector indicating a connection with stripe formation. However, figure 3 demonstrates that phonon renormalization is not simply a result of static stripe order as the effect at $\mathbf{q}_{co}$ is only slightly reduced after heating to a temperature just above the stripe ordering temperature of 60K. There is a slight suppression of the low energy tail and an enhancement of the peak, but the lineshape remains highly anomalous even when static stripes are no longer present.

In order to see whether a similar phonon anomaly is present in an optimally-doped superconductor, i.e. $La_{1.85}Sr_{0.15}CuO_4$, where there are no static stripes, we compare phonon spectra in the superconductor and in the stripe phase in Figure 4. The energy scans were taken at wavevectors $\mathbf{q}_{co}=(0.25,0,0)$ and $\mathbf{q}=(0.15,0,0)$, which are separated by slightly more than the full width at half maximum (FWHM) of the instrumental $\mathbf{Q}$-resolution. Whereas at $\mathbf{q}=(0.15,0,0)$ there is a single sharp peak, at $\mathbf{q}_{co}$ the main peak is much weaker and the missing intensity appears in the low energy tail in both compounds. This difference is strongest at the lowest temperature, 10K (Fig.3a,b). Obviously, it is somewhat stronger in the static stripe phase than in optimally doped $La_{1.85}Sr_{0.15}CuO_4$ but still very pronounced in the latter compound. On heating to T = 330K, the peak at $\mathbf{q}=(0.15,0,0)$ broadens considerably, presumably due to phonon-phonon interaction. The opposite happens at $\mathbf{q}_{co}$: the low-energy tail is greatly reduced and the main peak is enhanced. However, a low-energy tail is observed in $La_{1.875}Ba_{0.125}CuO_4$ even at 330K, whereas in $La_{1.85}Sr_{0.15}CuO_4$, the lineshape at $\mathbf{q}_{co}$ is conventional, albeit still rather broad. Further measurements revealed that in $La_{1.85}Sr_{0.15}CuO_4$ the maximum linewidth is not at **q**=(0.25, 0, 0) but at a slightly higher q value, i.e. **q**=(0.3,0,0) (see Fig. 7). A special attempt was made to see whether the lineshape at **q**=(0.3,0,0) changes on crossing the superconducting transition temperature of 35K. From inspection of Fig. 5a it is clear that this is not the case, although the peak does sharpen on heating to 330K (Fig. 5b).

In $La_{1.875}Ba_{0.125}CuO_4$ the anomaly rapidly disappears as one moves away from the (1 0 0) high symmetry line in the transverse [0 1 0] direction. Whereas at **Q**=(4.725 0 0) the phonon is extremely broad (Fig. 6a), a resolution-limited linewidth is recovered at **Q**=(4.725 0.15 0) (Fig. 6b). A constant-E scan at 65 meV in the transverse direction with **q**=(0.275 k 0), where -0.3<k<0.3, shows a peak with a width in **Q** that is nearly resolution-limited (Fig. 6c). Similar behavior is observed in $La_{1.85}Sr_{0.15}CuO_4$, although the transverse width appears to be larger than in $La_{1.875}Ba_{0.125}CuO_4$. (Fig. 6c). As will be discussed below, the narrow width of the phonon anomaly in the transverse direction suggests that it is not due to Fermi surface nesting (Kohn effect) but rather points to a coupling to a charge density wave modulation which is coherent in the transverse direction, such as charge stripes. In this sense, the somewhat larger transverse width in superconducting $La_{1.85}Sr_{0.15}CuO_4$ would mean a shorter coherence length.

The full spectrum of the bond-stretching vibration in $La_{1.875}Ba_{0.125}CuO_4$ can be fit with a superposition of two phonon branches of equal intensity dispersing as the solid lines in Fig. 2a. One branch has a monotonic, cosine-like downward dispersion in the (1 0 0) direction. In some sense, it can be considered as "normal" although it is well known that the corresponding branch is nearly flat in the undoped parent compound.[7] The other branch is highly anomalous in that it drops abruptly by more than 10 meV at about $q_{co}$ and then rises back up towards the zone boundary to merge with the normal branch. The drop is much sharper than the instrumental resolution in both the longitudinal and the transverse directions, so the average over the wave vectors within the resolution ellipsoid produces the broad low-energy tail at $Q_{co}$. No such branch splitting was observed in $La_{1.85}Sr_{0.15}CuO_4$ where it is always possible to fit the phonon spectra with a single peak, but the spectra show a large increase in linewidth as well as a sudden softening near q=(0.3 0 0) (Fig. 7).

We note that the data shown in Fig. 2 for $La_{1.875}Ba_{0.125}CuO_4$ were taken before we were aware of the narrow transverse width of the anomaly. For various reasons, we mounted the sample with the b-axis vertical. As a consequence of the relaxed q resolution of the neutron spectrometer in the vertical direction, this resulted in a relatively poor q resolution in the (010) direction. In our later measurements on $La_{1.875}Ba_{0.125}CuO_4$ and on $La_{1.85}Sr_{0.15}CuO_4$, the sample was mounted with the c-axis vertical, which improved the q-resolution in the (010) direction at the expense of the q-resolution in the (001) direction. A relatively poor q-resolution in the (001) direction is, however, irrelevant for these measurements because the phonon anomaly was found to be independent of the c-component. Thus, we do not have results for $La_{1.85}Sr_{0.15}CuO_4$ that can be compared quantitatively with Fig. 2. Instead, we use previously published data[6] for another stripe compound, i.e. $La_{1.48}Nd_{0.4}Sr_{0.12}CuO_4$, for a comparison with $La_{1.85}Sr_{0.15}CuO_4$, because they were taken with the same sample orientation (we note that our extensive measurements on both $La_{1.48}Nd_{0.4}Sr_{0.12}CuO_4$ and $La_{1.875}Ba_{0.125}CuO_4$ have shown that the two static stripe phases have very similar phonon properties). From Fig. 7 it is clear that the anomaly, characterized by a sharp drop in phonon frequency with increasing **q** and by broadening, is nearly as pronounced in $La_{1.85}Sr_{0.15}CuO_4$ as in $La_{1.48}Nd_{0.4}Sr_{0.12}CuO_4$ although the frequency drop is not quite as abrupt. The figure further shows that the anomaly shifts to somewhat higher q on increasing the Sr content from x = 0.12 to x = 0.15.

## 4. Discussion

The sharp drop in frequency and the large increase in linewidth at about **q** = (0.25,0,0) in $La_{1.85}Sr_{0.15}CuO_4$ demonstrated above are somewhat reminiscent of effects reported by McQueeney et al.,[8] and thus we believe that that study found manifestations of the same effect. However, our measurements do not confirm the claim that the bond-stretching branch splits into two nearly flat branches, and hence we find no evidence for the unit cell doubling at low temperatures advocated in that study. We emphasize that our measurements were performed with considerably better resolution and therefore, a branch splitting as reported in Ref. 8 should have been easily visible.

A very steep dispersion of the bond-stretching branch in $La_{1.85}Sr_{0.15}CuO_4$ in the (100)-direction half-way to the zone boundary was reported earlier by Pintschovius and Braden[9]. Moreover, they demonstrated that the phonon linewidths show a maximum around **q** = (0.3,0,0). These observations led them to conclude that $La_{1.85}Sr_{0.15}CuO_4$ is close to stripe phase order with a periodicity close to that observed in static stripe phase compounds. We note that the data presented in this paper for $La_{1.85}Sr_{0.15}CuO_4$ are largely consistent with those reported by Pintschovius and Braden[9]. However, our results show that the signatures for dynamic stripe order, i.e. a very steep dispersion near **q** = (0.25,0,0) and a concurrent maximum in the phonon linewidth, are considerably more pronounced than reported in ref. 9. This can be understood from the fact that in the measurements of ref. 10, the sample was oriented with a (010)-axis vertical whereas it was the (001)-axis in our study. As was discussed in the previous section, the relatively relaxed q-resolution of a neutron spectrometer in the vertical direction leads to a reduction of the stripe signatures if the sample is oriented with the (010)-axis vertical because these signatures rapidly disappear when going away from the (100)-direction in the (010)-direction. This was not known at the time of the previous study[9]. It was a major motivation for our study to repeat the previous measurements while avoiding the deleterious effects of the relaxed vertical resolution.

We now turn to the temperature dependence of the phonon linewidths displayed in Fig. 4. It is difficult to explain the phonon temperature dependence at **q** =(0.25,0,0) without invoking a tendency to a lattice instability, since conventional phonons should *broaden* with increasing temperature due to increased anharmonicity, not *sharpen* as we observe. The observed phenomenon does resemble a conventional charge density wave (CDW) instability in the sense that it points to a charge modulation with a characteristic wavevector. However, several independent electron-phonon coupling calculations sensitive to conventional CDW instabilities do not predict the huge effect that is observed.[10-13] They all give a very similar Fermi surface and adequately account for many nontrivial features of the phonon spectrum such as the downward dispersion of the normal bond-stretching branch and the strong electron-phonon coupling for the bond-buckling branch dispersing from the $B_{1g}$ Raman mode near 42 meV. On the other hand, these calculations greatly underestimate the renormalization of the bond-stretching branch at half-way to the zone boundary. For example, Refs. 10 and 11 give an electron-phonon coupling constant for the bond-stretching branch that is much smaller than that for the bond-buckling branch, which is renormalized by at most 1meV. (This can, for example, be read off from Fig. 1 in Ref. 10.) Furthermore, the narrow transverse **q**-width of the anomalous phonon (Fig. 6) is incompatible with the calculated[10,14] electron-phonon coupling in the transverse direction, which shows only gradual variations. (We note that we have similar difficulty when trying to obtain this effect from our shell model[7], which we have implemented by a special term to mimic the anomalous frequency drop at **q** = (0.25,0,0)). Finally, the apparent splitting of the bond-stretching branch into a normal and an anomalous one in the stripe phase is difficult to reconcile with a precursor effect to a conventional CDW: such a splitting is not expected in a system with tetragonal symmetry and indicates a local symmetry breaking.

It is apparent that one must go beyond a Fermi-liquid approach to find a possible explanation for our results. Becca *et al.*[15] considered a charge-density modulation close to an instability to phase separation. They found a very strong phonon softening at the wave vector of the charge instability, although many of the details differ significantly from our results. Alternatively, a relatively simple model assuming stripe order,[16] can explain most of the observed features in the phonon measurements. According to this model, collective electronic excitations emanating from the stripe ordering wavevector interact with phonons at the intersection of the electronic and the phonon dispersion curves. The strongest coupling is expected for bond-stretching modes, because the charge modulation gives rise to a bond length modulation. We note that this expectation is very well borne out by experiment: not only that we found a pronounced anomaly in the bond-stretching branch but further measurements also showed the lack of any discernable anomaly in other branches. Due to the steep dispersion of the electronic modes in both the transverse and the longitudinal directions, they interact with the phonons in a very limited range of reciprocal space. Thus this model explains the narrow **q**-width of the anomaly in both the transverse and longitudinal directions. The two-branch behavior also has a natural explanation: Stripes introduce a mechanism of local breaking of the tetragonal symmetry of the $CuO_2$ plane distinguishing between the phonons propagating parallel and perpendicular to the stripes. Since the crystals are macroscopically twinned with respect to the stripe direction, each branch will contribute equally to the neutron spectra resulting in the normal and the anomalous branch. However, it is not yet fully clear whether there is a simpler explanation for the apparent two-branch behavior, i.e. that it is an artifact of finite q-resolution in conjunction with the narrow width of the anomaly in the transverse direction. We are currently investigating this possibility.

M. Vojta et al.[17] have shown that such symmetry breaking may occur only very locally when stripes are dynamic, since stripe fluctuations may occur in both directions simultaneously. Consistent with this picture, the phonon lineshape in $La_{1.85}Sr_{0.15}CuO_4$ can be well fit with a single broad peak.

Of course, in considering the nature of the stripe correlations and the associated electron-phonon interactions, one must keep in mind that there are further constraints from electron spectroscopies. It has recently been shown by optical conductivity[18] and angle-resolved photoemission[19] studies that the ordered stripe phase is not insulating, at least not in any conventional sense. In fact, it exhibits a *d*-wave-like gap. Thus, the stripe phase is not compatible with the conventional concept of a charge-density wave.

## 5. Conclusions

As discussed above, there is good reason to attribute the giant renormalization in the bond-stretching phonon branch near **q**=(0.25 0 0) in $La_{1.48}Nd_{0.4}Sr_{0.12}CuO_4$ and $La_{1.875}Ba_{0.125}CuO_4$, which have low temperature static stripe order, to a coupling of the phonons to charge density wave formation in the form of stripes. Therefore, we consider the fact that a similar feature is observed in optimally doped $La_{1.85}Sr_{0.15}CuO_4$ as evidence that superconducting samples with a high $T_c$ are close to charge stripe formation as well. In the superconducting samples, however, the stripes never become static but remain dynamic. Apparently, a strong tendency towards stripe formation is not restricted to a narrow range of the phase diagram close to the so-called 1/8-anomaly but is observed for optimally doped samples as well. On the other hand, recent investigations of heavily overdoped, nonsuperconducting $La_{1.7}Sr_{0.3}CuO_4$ by inelastic x-ray[20] and neutron[21] scattering did not show any signatures of dynamic charge stripes. This indicates that stripe formation is correlated with the occurrence of high $T_c$ superconductivity and therefore might be important for its understanding.

# 6. Acknowledgments

Work at Brookhaven is supported by the Office of Science, U.S. Department of Energy, under Contract No. DE-AC02-98CH10886.

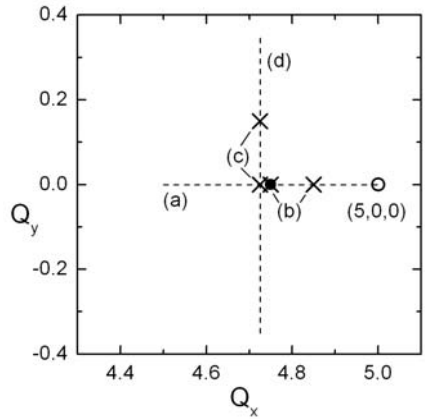

Figure 1 Reciprocal space positions where the scans were made. The black dot represents the stripe-ordering wavevector. a) scans in Fig. 2a; b) scans in Figs. 3, 4 and 5; c) scans in Fig. 6a,b; d) scan in Fig. 6c at 65 meV.

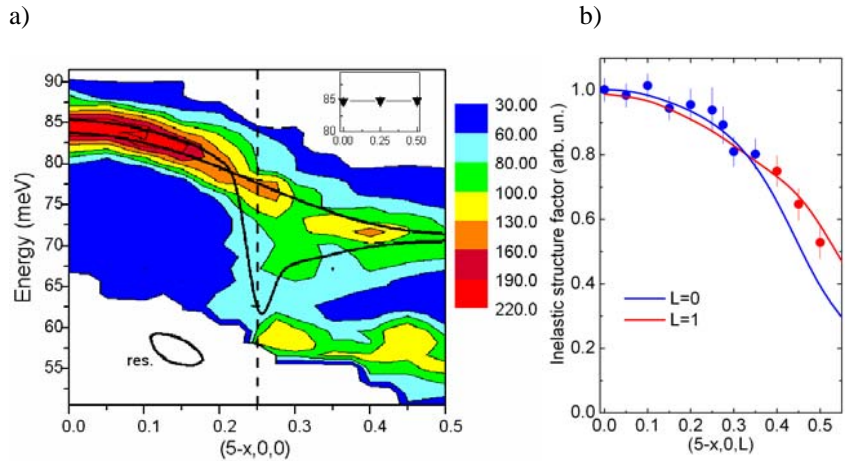

Figure 2 a) Color-coded contour plot of the intensities observed on $La_{1.875}Ba_{0.125}CuO_4$ at T = 10 K. (From Ref. 6) The intensities above and below 60 meV are associated with plane-polarized Cu-O bond-stretching vibrations and bond-bending vibrations, respectively. Black lines are dispersion curves evaluated from two-peak fits to the data (one of them has a cosine-behavior). The white area at the lower left corner of the diagram was not accessible in this experiment. The ellipse illustrates the instrumental resolution. The inset shows the dispersion in the 110-direction. The dashed line represents the charge–ordering wavevector.

b) The dots denote inelastic structure factors of the bond-stretching modes as deduced from the integrated intensities observed for wave vectors **Q** = (5-x,0,0) (blue) and **Q** = (5-x,0,l) (red). The data were normalized to the value at **Q** = (5,0,0). The solid lines were calculated from a shell model.[7] We note that the inelastic structure factors depend very little on the value of L except for wave vectors close to the in-plane zone boundary.

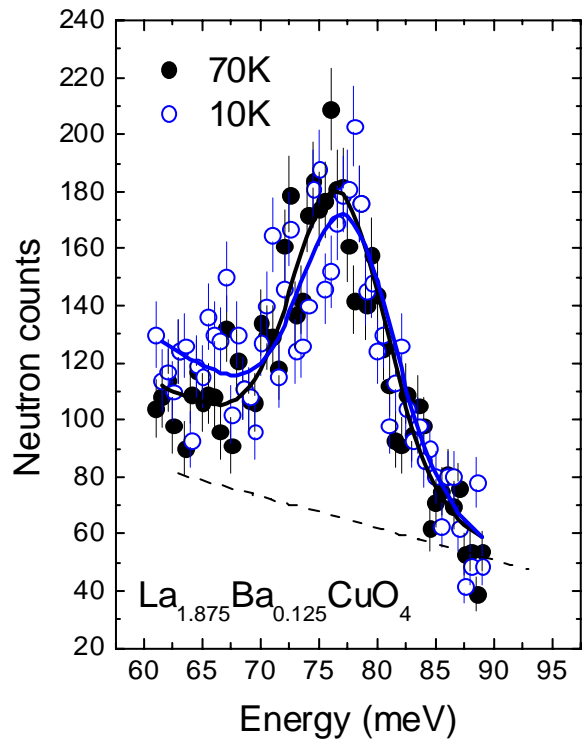

Figure 3 Anomalous phonon at **q**=(0.25 0 0) in $La_{1.875}Ba_{0.125}CuO_4$ above and below the stripe-ordering temperature T = 60 K. Lines through the data are guides for the eye. Straight dashed line represents the background.

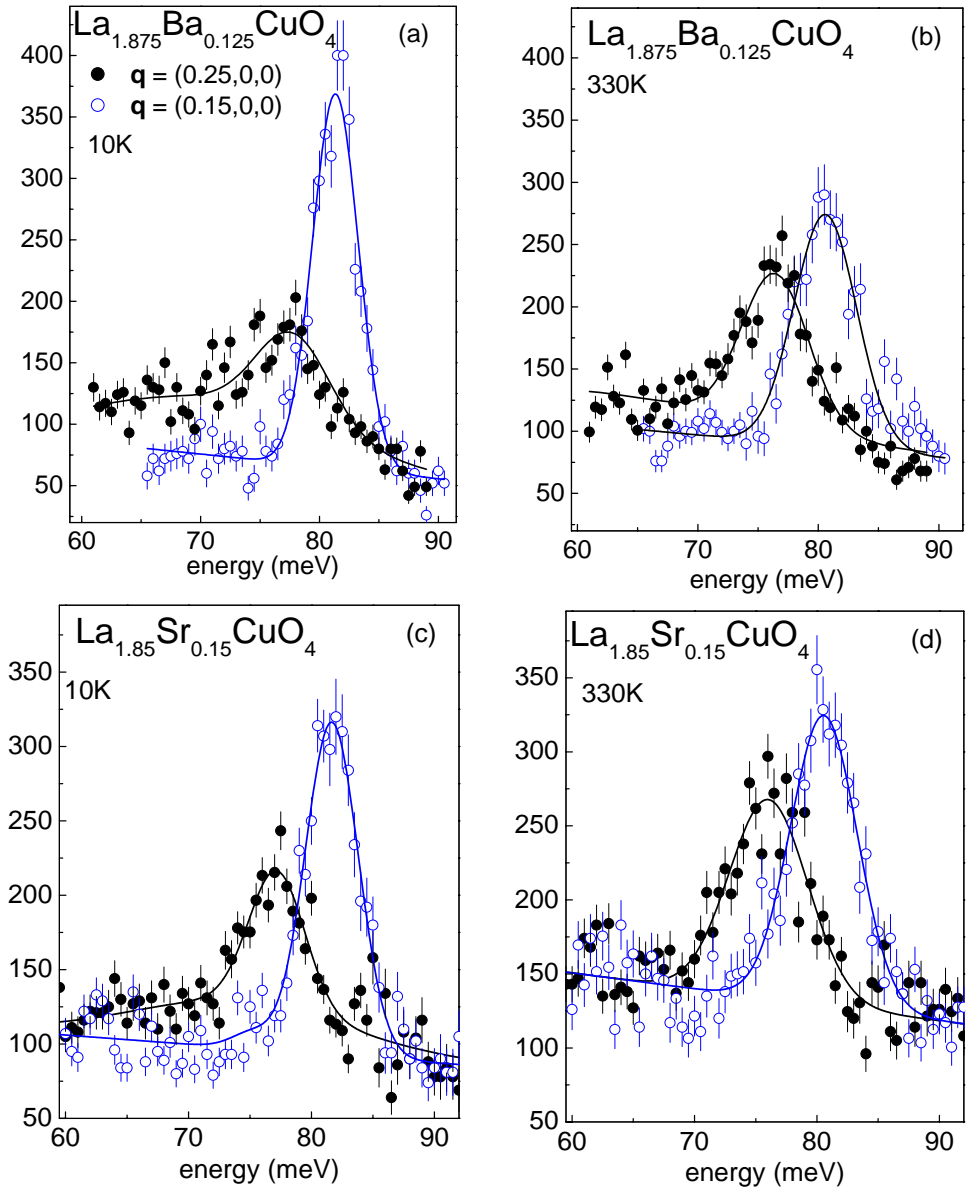

Figure 4 Energy scans taken on $La_{1.875}Ba_{0.125}CuO_4$ (a,b) (Ref. 6) and on $La_{1.85}Sr_{0.15}CuO_4$ (c,d) at 10K (a,c) and 330K (b,d). The phonon at $\mathbf{Q}=(4.85\ 0\ 0)$ is "normal" in that it has a Gaussian lineshape on top of a linear background. This background results from multiphonon and incoherent scattering and has no strong dependence on $\mathbf{Q}$. The intensity reduction of this phonon in $La_{1.875}Ba_{0.125}CuO_4$ from 10K (b) to 330K (c) is consistent with the Debye-Waller factor. At $\mathbf{Q}=(4.75\ 0\ 0)$, there is extra intensity on top of the background in the tail of the main peak. It results from one-phonon scattering that extends to the lowest investigated energies, while the peak intensity is greatly suppressed as discussed in the text. The effect is reduced but does not disappear

at 330K. Note that 330K in (b) is shown instead of 300K in the same plot in Ref. 6 is because of a typographical error in the latter.

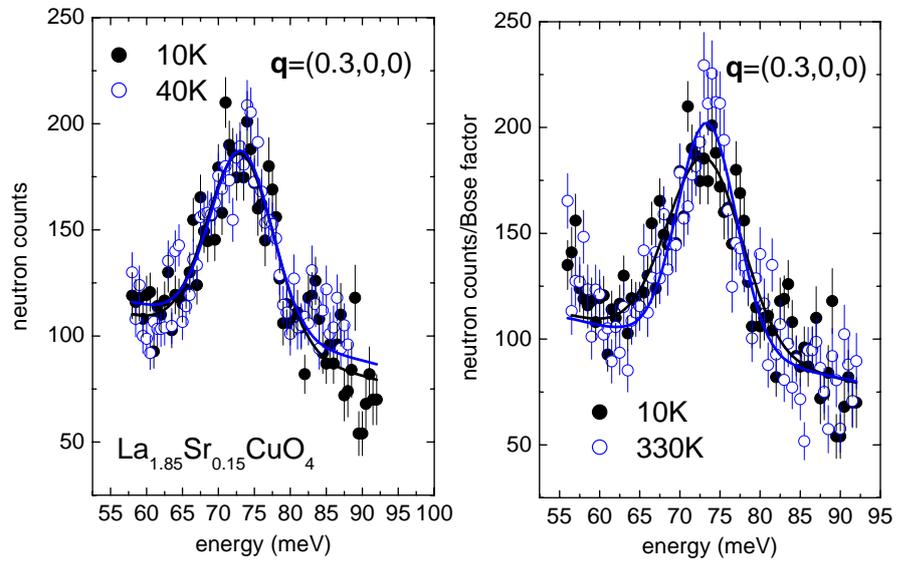

Figure 5. Temperature dependence of the anomalous phonon at Q=(4.7 0 -2) in $La_{1.85}Sr_{0.15}CuO_4$. Lines are guide for the eye. The 330K spectrum was shifted down by 30 counts to compensate for increased background.

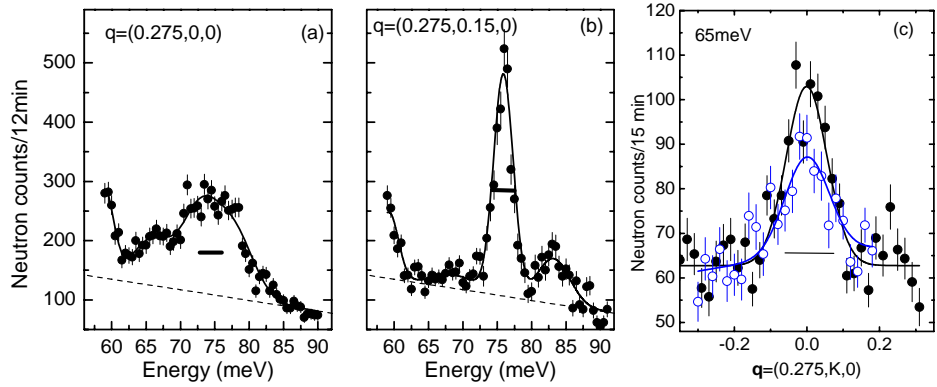

Figure. 6 (a,b). Energy scans taken on $La_{1.875}Ba_{0.125}CuO_4$ at wave vectors (a) $Q = (4.725,0,0)$ and (b) $Q = (4.725,0.15,0)$, respectively. The horizontal bars denote the instrumental resolution. The dashed lines denote the background. Note that in (a) where $Q = (4.725,0,0)$, the selection rules are such that only longitudinal modes contribute. In (b) where $Q = (4.725,0.15,0)$, the symmetry is lower, and modes of transverse character become allowed as well. This results in two peaks appearing instead of one (at 75 and 83meV). Their character is neither purely longitudinal nor transverse but mixed. Longitudinal character dominates in the strong lower energy peak, whereas the weaker higher energy peak is mostly transverse. Their relative contribution is in accordance with predictions from a lattice dynamic model.

(c). $Q$-scan in the transverse direction, i.e. in the (010)-direction, through the low energy tail (E = 65 meV) of the bond-stretching phonon peak (maximum at E = 75 meV) centered at $Q = (4.725\ 0\ 0)$. The line is a Gaussian fit. Solid circles correspond to $La_{1.875}Ba_{0.125}CuO_4$ and the open ones to $La_{1.85}Sr_{0.15}CuO_4$ Horizontal bar denotes the instrumental resolution. The narrow intrinsic peak width indicates that the anomalous behaviour is restricted to phonons propagating along the 100-direction.

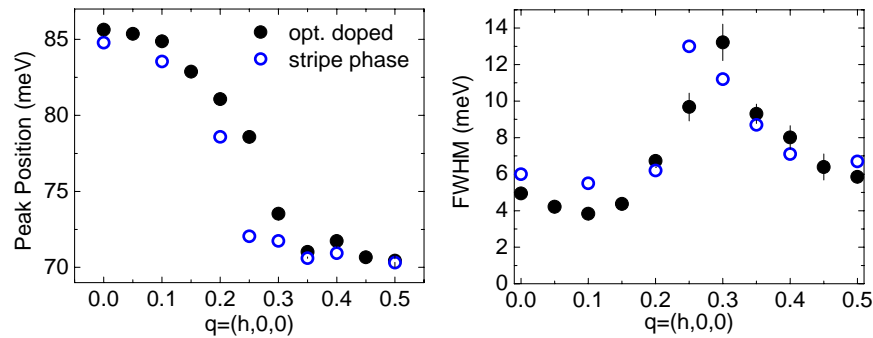

Figure 7  Comparison of the phonon dispersions (a) and linewidth (b) of the bond-stretching branch in $La_{1.85}Sr_{0.15}CuO_4$ and $La_{1.48}Nd_{0.4}Sr_{0.12}CuO_4$.

14